\documentclass[aps,prl,twocolumn,10pt,floatfix]{revtex4-1}

\usepackage[OT1]{fontenc}
\usepackage[utf8]{inputenc}
\usepackage{graphicx}
\usepackage{braket}
\usepackage{siunitx}
\usepackage{amsmath}
\usepackage{amssymb}
\usepackage{bm}
\usepackage{booktabs}
\usepackage[capitalize]{cleveref}

\usepackage{color}

\begin{document}

\title{Conformational Dynamics Guides Coherent Exciton Migration in Conjugated Polymer Materials: A First-Principles Quantum Dynamical Study}

\author{Robert Binder,$^\dagger$ David Lauvergnat,$^\ddag$ and Irene Burghardt$^{\dagger}$}
\email[Author to whom correspondence should be addressed: ]{burghardt@chemie.uni-frankfurt.de}
\affiliation{$^\dagger$Institute of Physical and Theoretical Chemistry, Goethe University Frankfurt,
             Max-von-Laue-Str.\ 7, 60438 Frankfurt, Germany \\
             $^\ddag$Laboratoire de Chimie Physique, Universit{\'e} Paris-Sud, UMR 8000, 91405 Orsay, France}

\date{\today}

\begin{abstract}
We report on high-dimensional quantum dynamical simulations of torsion-induced
exciton migration in a single-chain oligothiophene segment comprising twenty
repeat units, using a first-principles parametrized Frenkel $J$-aggregate
Hamiltonian. Starting from an initial inter-ring torsional defect, these simulations
provide evidence of an ultrafast two-time scale process at low temperatures,
involving exciton-polaron formation within tens of femtoseconds, followed by
torsional relaxation on a $\sim$300 femtosecond time scale. The second step is the
driving force for exciton migration, as initial conjugation breaks are removed
by dynamical planarization. The quantum coherent nature of the elementary
exciton migration step is consistent with experimental observations
highlighting the correlated and vibrationally coherent nature of the dynamics
on ultrafast time scales.
\end{abstract}

\maketitle

The photogeneration and spatial migration of excitons is of key
importance in organic electronic materials \cite{KB2015}. 
Molecular excitons in these $\pi$-conjugated organic semiconductors are bound electron-hole pairs delocalized over
one to ten monomer units \cite{Hildner2016,Grage03,Tretiak02}, whose diffusion length typically falls
into a 10 nm range. The basic entities subject to photoexcitation
have also been termed chromophores, or spectroscopic units \cite{Grage03,Beenken04}, and have been 
identified with 
conformational subunits delimited by conjugation breaks or
chemical defects. However, the precise definition of spectroscopic units
and their dynamical evolution, at a molecular level, has remained elusive \cite{Beenken04,Beenken09}.

While conventional ensemble measurements have been limited by
conformational averaging, recent single-molecule
low-temperature spectroscopy \cite{Hildner2016} as well as
transient electronic spectroscopies \cite{Scholes09} and time-resolved
Raman studies \cite{Bragg16} pave the way for a more detailed
understanding of the primary excitations and their dynamics, on time 
scales ranging from femtoseconds to hundreds of picoseconds. Various studies
point to the dominant role of conformational defects along polymer chains,
and emphasize that fluctuations from disordered to ordered conformations are
favored on short time scales.
Torsional relaxation has been
conjectured to interfere with excitation energy transfer \cite{Sundstrom06}, and 
the large Stokes shift is indicative of strong
exciton-phonon coupling leading to highly correlated motions in a
subpicosecond regime \cite{Blank2008}.
Vibrational quantum coherence was found to persist
on time scales of several hundred femtoseconds \cite{Song2015}.
Meanwhile, the role of exciton trapping, exciton-polaron formation, and the response of excitonic species to structural
changes remain controversial \cite{Bragg16,Barford17}.

Theoretical analysis of exciton transport in single chains (i.e., quasi-1D systems)
generally relies on a Frenkel-Holstein Hamiltonian \cite{Holstein59} description for $J$-aggregates, i.e.,
head-to-tail aligned molecular aggregates \cite{Spano14}. 
Besides F\"orster rate theory \cite{KB2015}, quantum-classical studies of
Ehrenfest type \cite{Barford12a} and Surface-Hopping type \cite{Nelson2017}
have been conducted, all of which
are restricted in the treatment of exciton-phonon correlations
and may not correctly describe the intricate interplay of electronic
delocalization, trapping, and exciton migration. Full quantum dynamical
studies
have been limited to short oligomer segments
where the earliest time scales were
found to be dominated by coherent electronic
effects \cite{Faraday2013}.

\begin{figure}[b]
  \includegraphics[width=1\columnwidth]{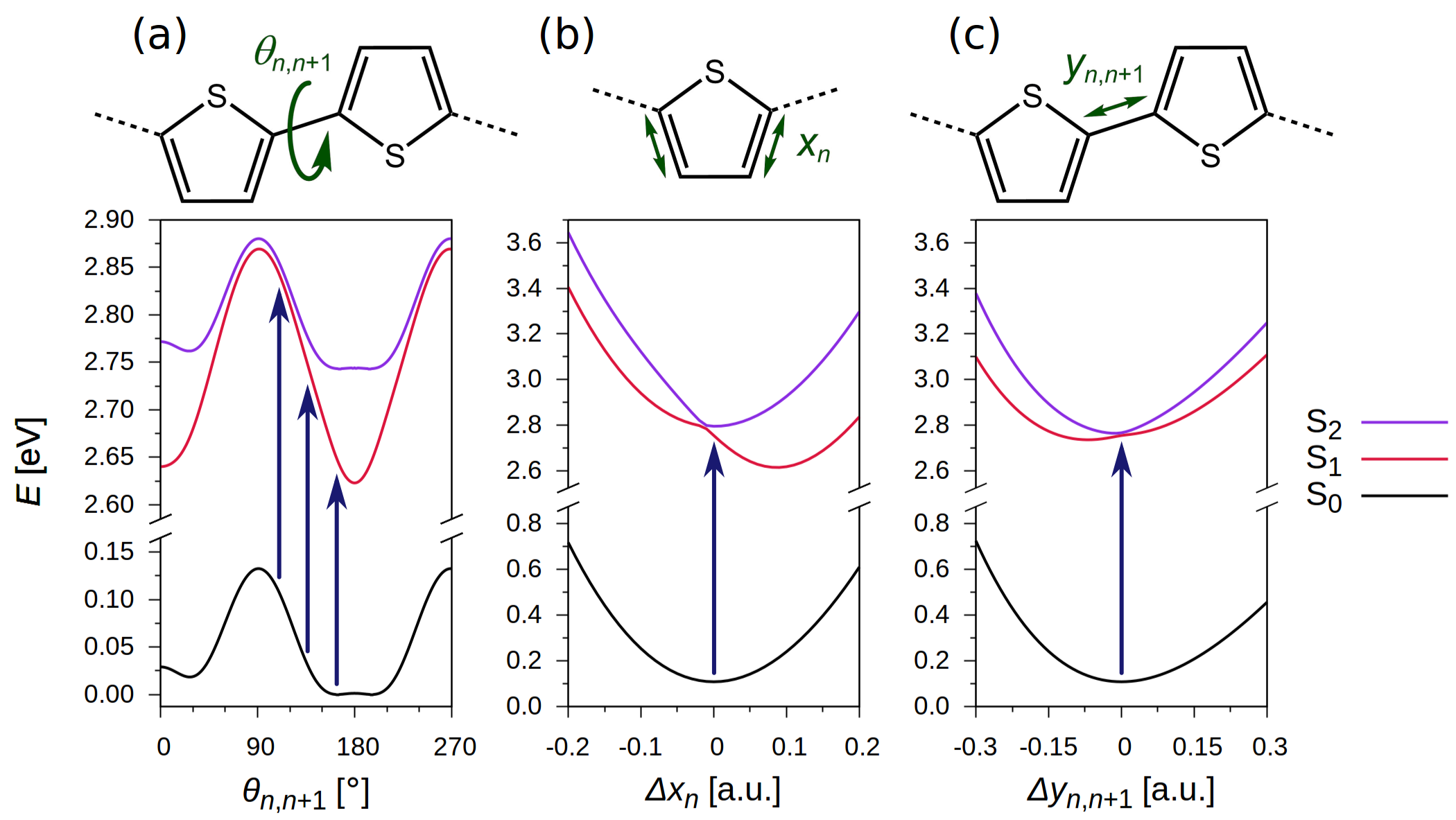}
  \caption{{\em Ab initio} based PES sections for an OT-20 oligomer as a
    function of selected torsional ($\theta_{n,n+1}$), ring-breathing
    ($x_{n}$), and inter-monomer bond stretch ($y_{n,n+1}$) coordinates.
  The electronic ground state ($S_0$) and first two excited singlet states
  ($S_1$, $S_2$) are shown. Note that the torsional $S_1$ potential is
  very stiff as compared with the ground state; also, the local ring-breathing
  mode exhibits the largest PES shift, giving rise to
  pronounced exciton trapping effects.}
\label{figure1}   
\end{figure}

Against this background, the present study employs a
high-dimensional quantum dynamical description, using
efficient multiconfigurational approaches \cite{mctdh2000,Wang2015}, to elucidate the
elementary step of intra-chain exciton migration in the presence of torsional dynamics.
We employ an {\em ab initio} parametrized, generalized
$J$-aggregate Hamiltonian for an
oligothiophene eikosamer (OT-20)
representative of a chromophore segment of poly(3-hexylthiophene) (P3HT) which plays
a central role as an electron donor species in organic photovoltaics. 
The Hamiltonian was constructed by a
mapping procedure \cite{Binder2014} which transforms oligomer
potential energy surfaces (PES) to a site-based Frenkel-Holstein type
Hamiltonian. Selected high-level electronic
structure calculations were carried out to construct PES sections for
several modes exhibiting the most pronounced vibronic coupling effects,
i.e., inter-monomer torsional modes ($\theta_{n,n+1}$),
symmetric ring-breathing modes ($x_{n}$), and  
inter-monomer C-C bond stretch modes ($y_{n,n+1}$), as illustrated in Fig.\
\ref{figure1}. 

The resulting Hamiltonian in the basis of Frenkel configurations $\ket{n}$, i.e.,
single-monomer excitations in the direct-product space spanned by $N$
monomers, reads
\begin{eqnarray}
  \label{H_Frenkel}
  \hat{H} & = & \sum_{n,n'=1}^N \hat{H}_{n,n'} \ket{n}\bra{n'} + \hat{H}_{\rm bath} \hat{1}
\end{eqnarray}
combining contributions for an $N$-site system ($\hat{H}_{n,n'}$) with an external bath ($\hat{H}_{\rm bath}$). Here,   
$\hat{H}_{n,n'}$ comprises kinetic energy ($\hat{T}$) and on-site ($\hat{V}_{n}^{\rm site}$) contributions, as well as
excitonic couplings ($\hat{V}^{\rm exc}_{n,n'}$), 
\begin{eqnarray}
 \hat{H}_{n,n'} & = & \delta_{n,n'} \hat{T} + \delta_{n,n'} \hat{V}_{n}^{\rm site} +
   \hat{V}^{\rm exc}_{n,n'}
\label{H_all}
\end{eqnarray}
where $\hat{T}$ corresponds to the kinetic energy operator in curvilinear coordinates \cite{Lauvergnat2002,SI},
\begin{eqnarray}
  \hat{T} & = &  
  \frac{1}{2} \biggl( \sum_{n=1}^{N} G_{xx} \hat{p}_{x_n}^2 + \sum_{n=1}^{N-1} ( G_{yy} \hat{p}_{y_{n,n+1}}^2 + G_{\theta\theta} \hat{p}_{\theta_{n,n+1}}^2) \biggr. \nonumber  \\
  \mbox{} & & \biggl. + 2 \sum_{n=1}^{N} G_{xy} \hat{p}_{x_n} ( \hat{p}_{y_{n,n+1}} + \hat{p}_{y_{n,n-1}}) \biggr)
\end{eqnarray}
with $G_{\theta\theta} = I_\theta^{-1}$, $G_{yy} = m_y^{-1}$, and $G_{xx} = m_x^{-1}$. The site potential of Eq.\ (\ref{H_all}) reads as follows, 
\begin{eqnarray}
    \hat{V}_{n}^{\rm site} ( \{ \hat{x}, \hat{y}, \hat{\theta} \} ) & = & \hat{V}_0( \{ \hat{x}, \hat{y}, \hat{\theta} \} ) + \hat{\Delta}_{n} ( \hat{x}_n, \hat{y}_{n,n\pm 1}, \hat{\theta}_{n, n\pm 1} )
    \nonumber \\
\end{eqnarray}
with the ground-state potential 
\begin{eqnarray}
    \hat{V}_0( \{ \hat{x}, \hat{y}, \hat{\theta} \} ) =  \sum_{l=1}^N \hat{v}_{\text{G}} \left( \hat{x}_l, \hat{y}_{l,l\pm 1}, \hat{\theta}_{l,l\pm 1} \right)
\end{eqnarray}
and the difference potential   
\begin{eqnarray}
  \hat{\Delta}_{n} ( \hat{x}_n, \hat{y}_{n,n\pm 1}, \hat{\theta}_{n, n\pm 1} ) & = & c_{\text{E}} + \hat{v}_{\text{E}} \left( \hat{x}_n, \hat{y}_{n,n\pm 1}, \hat{\theta}_{n,n\pm 1}
      \right) \nonumber \\
      \mbox{} & & \biggl. - \hat{v}_{\text{G}} \left( \hat{x}_n, \hat{y}_{n,n\pm 1}, \hat{\theta}_{n,n\pm 1} \right) 
\end{eqnarray}            
where $c_{\text{E}}$ is a constant excitation energy and 
$\hat{v}_{\text{G}}$ and $\hat{v}_{\text{E}}$ are ground-state and excited-state monomer potentials. Further, the excitonic
couplings of Eq.\ (\ref{H_all}) are given as
\begin{eqnarray}
    \hat{V}_{n,n\pm 1}^{\rm exc} (\hat{\theta}_{n,n\pm 1})  
      & = & \hat{w}(\hat{\theta}_{n,n\pm 1}) 
\end{eqnarray}
where an additional dependence on the high-frequency modes $\hat{x}_n$ and $\hat{y}_{n,n\pm 1}$ has been neglected.
The functions $v_{\text{G}}$ and $v_{\text{E}}$ (including edge effects \cite{SI}), along with $w$ and the
constant $c_{\text{E}}$ are determined by the solution of an inverse eigenvalue
problem as described in Ref.\ \cite{Binder2014}. In the present case, the functional forms correspond to
Morse potentials and a cosine series, respectively \cite{Binder2014,SI}.

\begin{figure}[b]
  \includegraphics[width=\columnwidth]{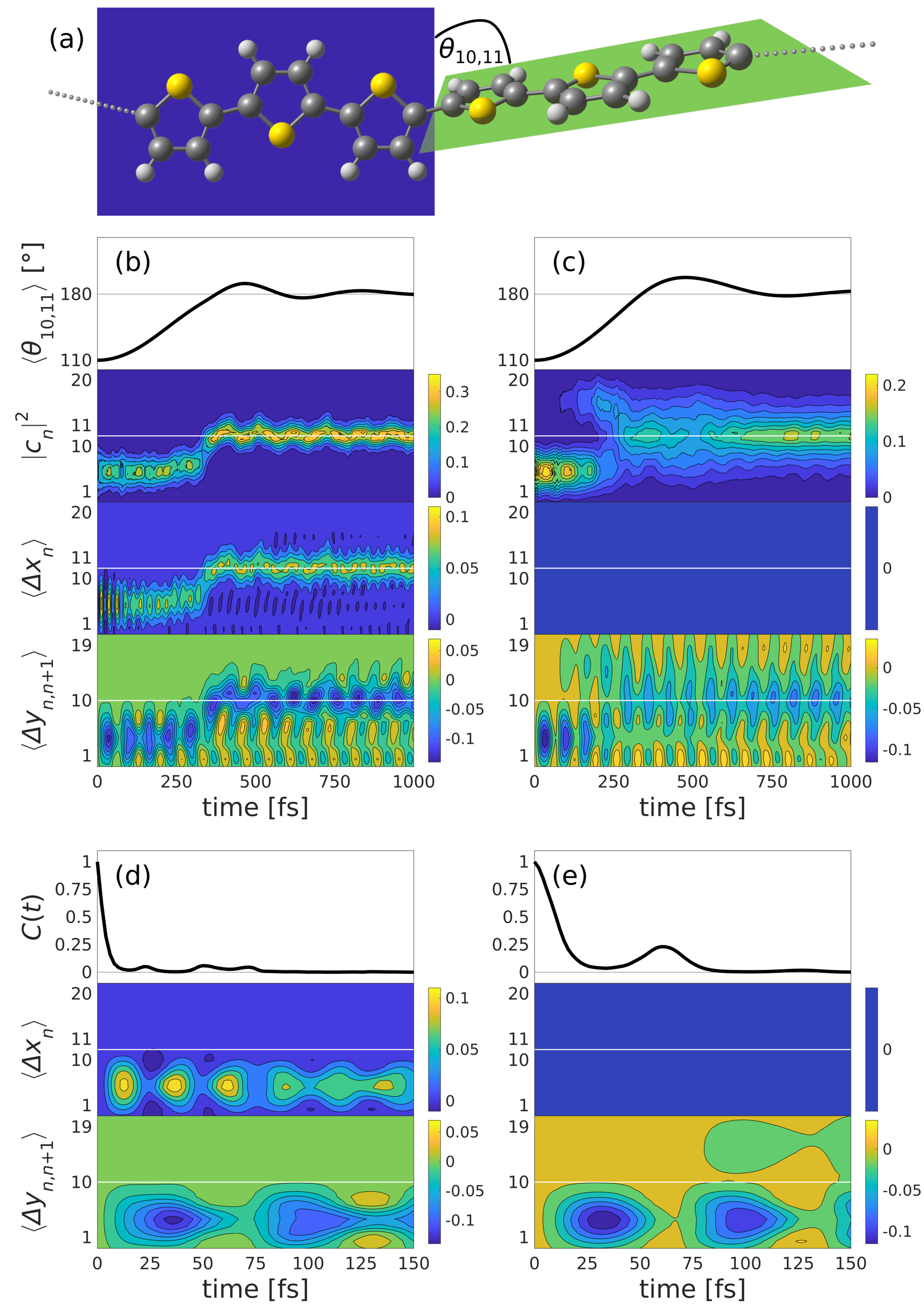}
  \caption{(a) Schematic illustration of the OT oligomer system under investigation, (b) for a simulation set-up including all modes, expection value of the central torsion
    ($\langle \theta \rangle_{10,11}$), squared coefficients of the
    excitonic wavefunction ($\vert c_n \vert^2$), expectation values
    of the high-frequency modes relative to their equilibrium
    position ($\langle \Delta x_n \rangle =
    \langle \hat{x}_n \rangle  -
    \langle \hat{x}_n \rangle_{0} )$ and analogously for
    $\langle \Delta y_{n,n+1} \rangle$), (c) analogous results for a simulation set-up
  excluding the local modes $\{ \hat{x}_n \}$, (d) on a shorter time scale (150 fs), the decay of the transition dipole autocorrelation function
  $C(t)$ is shown, along with expectation values of the high-frequency modes, (e) analogously, $C(t)$ in the absence of the local modes $\{ \hat{x}_n \}$.}
\label{figure2}  
\end{figure}      

Finally, the bath Hamiltonian of Eq.\ (1) corresponds to a collection of harmonic oscillators
coupled to the torsional modes $\theta$,
\begin{eqnarray}
  \hat{H}_{\rm bath} ( \{ \hat{b} \} ) & = & 
\sum\limits_{j} \frac{\hat{p}_{b,j}^2}{2}
+ \frac{\omega_{j}^2}{2} ( \hat{b}_{j} -
\frac{c_{j}}{\omega_{j}^2} ( \hat{\theta}_{n,n+1} - \theta_{{0}} ))^2  \quad
\label{HO-bath}
\end{eqnarray}
where the couplings $c_{j}$ are adapted to an Ohmic spectral density,
$c_{j} = \omega_{j} \sqrt{(2/\pi)\gamma I_{\theta} \Delta \omega}$
with $\gamma$ the friction coefficient and 
$\Delta \omega$ the frequency spacing that determines the Poincar\'e
recurrence time $T_P = 2 \pi/\Delta\omega$.

Quantum dynamical simulations were performed using
the Multi-Layer Multiconfiguration
Time-Dependent Hartree (ML-MCTDH) method \cite{mctdh2000,Wang2015}
(Heidelberg MCTDH package \cite{MCTDHHD}),
comprising
eight layers encompassing 50 phonon degrees of freedom and 20 electronic
states \cite{SI}.
To reduce the
computational effort, we only include the central inter-monomer torsion of the OT-20
segment, i.e., $\theta_{10,11}$, with an initial displacement at $\theta_{10,11}=\ang{110}$, 
and keep all other inter-monomer torsions fixed at the ground-state equilibrium
geometry, $\theta_{\text{eq}} = {160}^\circ$.
Damping is imposed on $\theta_{10,11}$ using the harmonic oscillator bath
of Eq.\ (\ref{HO-bath}) comprising
a set of at least 10 bath oscillators with a frequency spacing of $\Delta\omega=\num{6e-5}$
a.u.\ and a damping time of $\gamma^{-1}=\SI{200}{\femto\second}$.

\begin{figure}[b]
  \includegraphics[width=0.7\columnwidth]{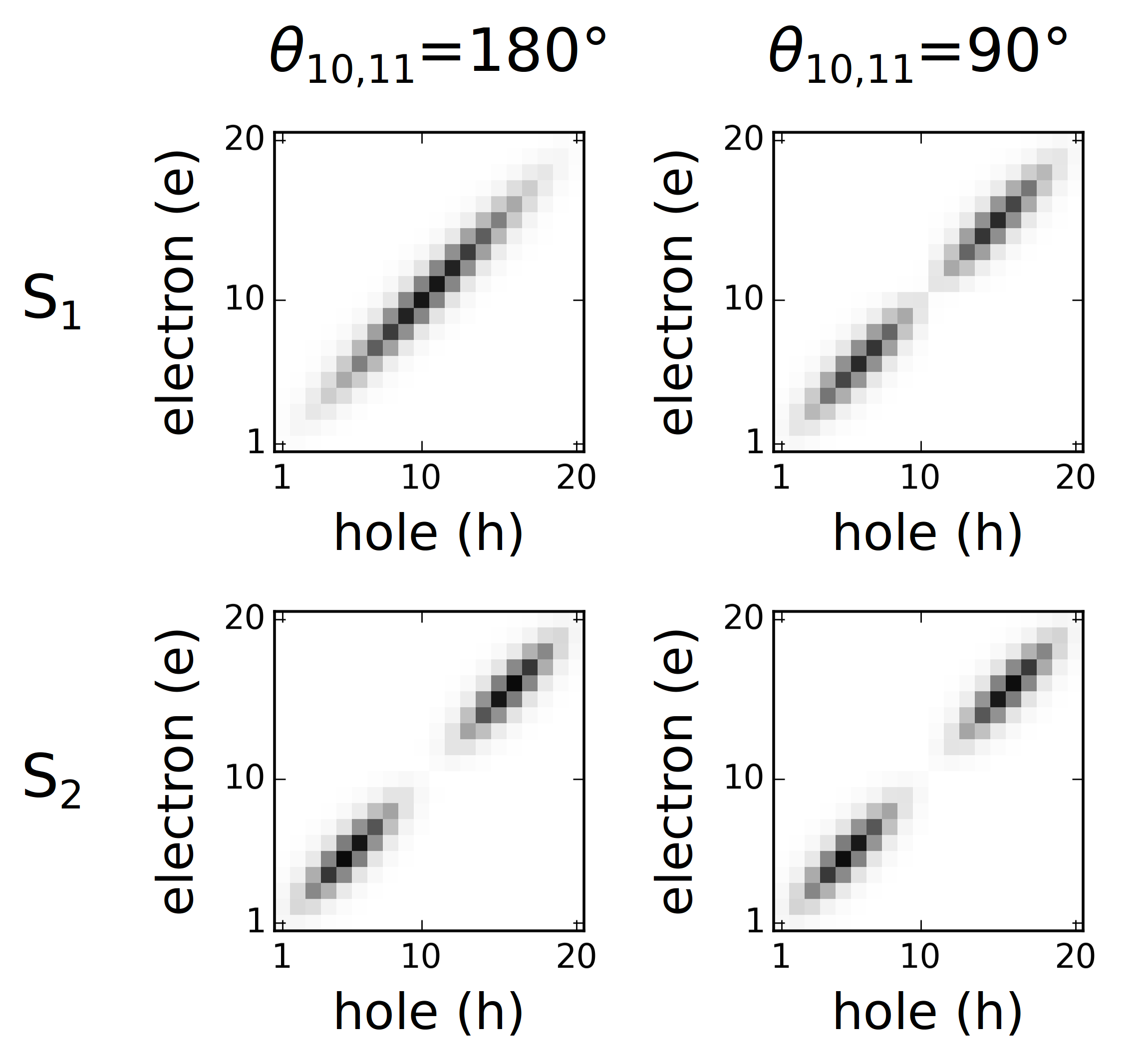}
  \caption{Electron-hole map, showing occupation
    probabilities of $\vert n_e n_h \rangle$ states, 
    obtained by TDA
  for OT-20, based on TDDFT calculations. The TDA analysis is
  shown for the two lowest adiabatic states ($S_1$, $S_2$), for the planar geometry ($\theta_{10,11} = 180^{\circ}$) and the
  twisted geometry ($\theta_{10,11} = 90^{\circ}$).}
  \label{figure3} 
\end{figure}      

Prior to real-time calculations, imaginary-time propagation in
the electronic subspace was carried out in order to create the initial,
partially delocalized exciton state to the left of the torsional defect,
which corresponds to the lowest adiabatic eigenstate of the left OT-10 segment.

The exciton evolution up to 1 picosecond, along with the expectation values of
the central torsional mode $\theta_{10,11}$ and the high-frequency modes, as
well as the transition dipole autocorrelation function
$C(t) = {\rm Tr} \{ \hat{\mu}(t) \hat{\mu} \ket{0} \bra{0} \} = \mu_{\text{EG}}^2 
\langle \psi (t) \psi(0) \rangle$ \cite{SI} are shown in Fig.\ \ref{figure2}. The excitonic
wavefunction is found to remain localized on the left segment during about 300
femtoseconds and is then displaced toward the center of the lattice, as the
central torsion planarizes. More detailed analysis shows that the interaction
of the exciton with the high-frequency vs.\ low-frequency modes gives rise to two characteristic
time scales, as now discussed.

Within the first $\sim$50 femtoseconds, the exciton contracts by three monomer units \cite{SI}, while the  
intra-monomer CC bonds ($x_n$) of the left fragment are found to expand and the inter-monomer CC bonds ($y_{n,n\pm 1}$) contract.
The concomitant decay of the transition dipole autocorrelation function $C(t)$
within $\sim$10 fs (Fig.\ \ref{figure2}d) 
can be related to the experimentally observed polarization
anisotropy decay \cite{Grage03b}. During the whole simulation, the mode displacements 
appear to move adiabatically with the exciton -- i.e., the exciton is ``dressed'' by the lattice
distortion, forming an exciton-polaron quasi-particle.

Between \SI{300}{\femto\second}
and \SI{400}{\femto\second}, the exciton-polaron is completely transferred to the center
of the chain (Fig.\ 2b). The transfer is rather sudden and
coincides with the dynamical planarization of the active torsional mode
$\theta_{10,11}$, in excellent agreement with the experimentally observed torsional relaxation
time scale of 400 fs \cite{Bragg15}.
The exciton is now trapped at the center of the chain, 
while the intra and inter-monomer CC bonds of the left
segment return to their ground state equilibrium values.
In the absence of the high-frequency local CC modes (Fig.\ 2c), the trapping effect
turns out to be conspicuously less pronounced, and the exciton density appears less compact.

\begin{figure}[b]
  \includegraphics[width=1\columnwidth]{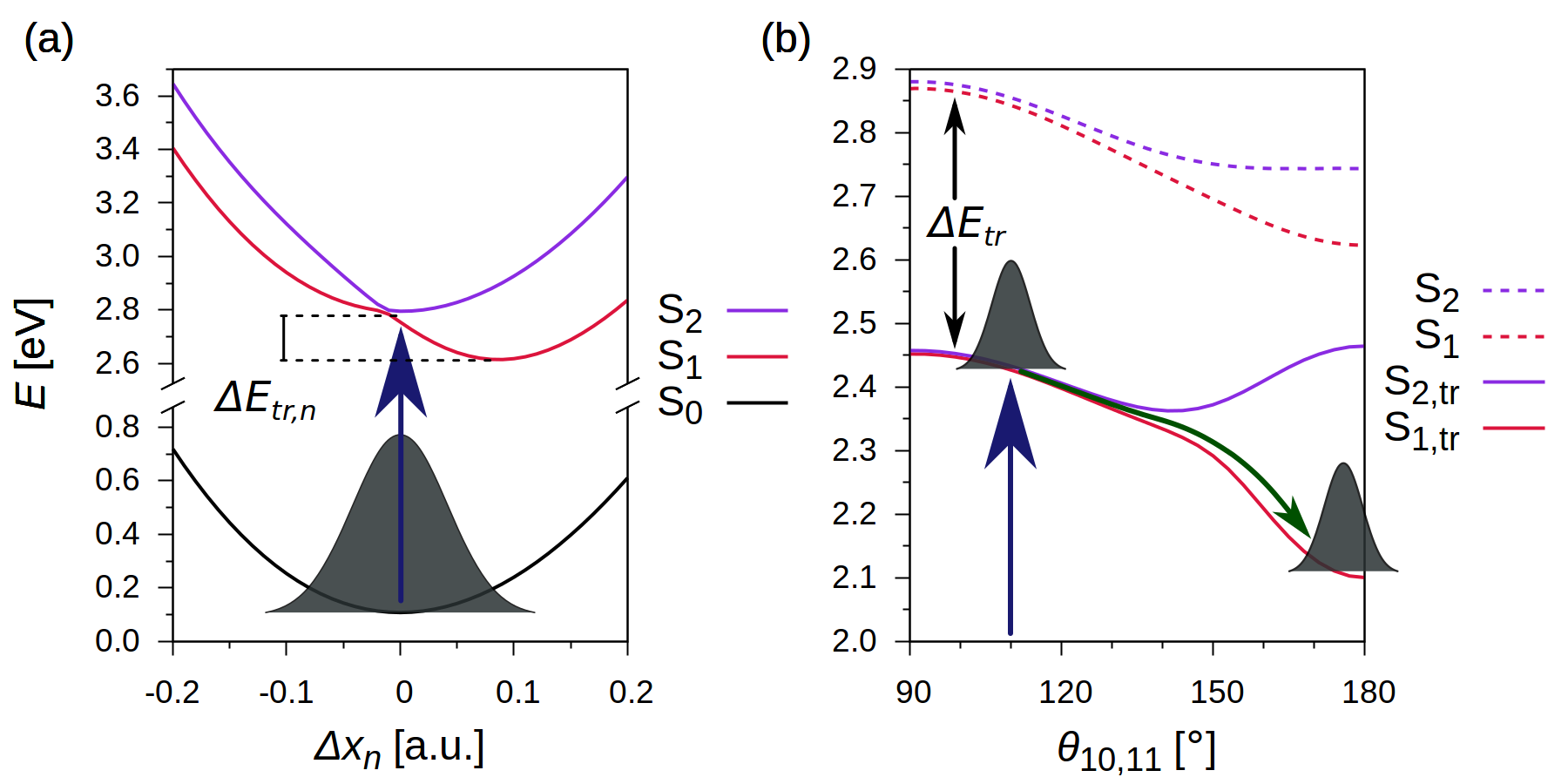}
  \caption{Main steps of the photoinduced process on coupled
    $(S_1, S_2)$ potentials:
  (a) Excited-state reorganization of local high-frequency modes
  $\{ \hat{x}_n \}$ with reorganization (trapping)  
    energy $\Delta E_{{\rm tr},n}$, (b) torsional relaxation on
    effective, dressed potentials ($S_{1,{\rm tr}}, S_{2,{\rm tr}}$) 
    including polaronic stabilization $\Delta E_{{\rm tr}}$.}
\label{figure4}
\end{figure}      

The role of the torsional degree of freedom is best appreciated by analyzing
the dependence of the electronic structure on the central torsion
$\theta_{10,11}$, see Fig.\ \ref{figure3}. Here, a transition density analysis (TDA) \cite{TM02,Panda2013} is
shown for OT-20, yielding an electron-hole representation $\vert n_e n_h
\rangle$, based upon Time-Dependent Density Functional Theory (TDDFT) calculations \cite{SI}. The diagonal dominance of the TDA
map indicates that the Frenkel model employed in the present study,
which is restricted to the center-of-mass exciton ($n_e=n_h$), is a good
approximation. As can be inferred from the TDA, the lowest excited electronic
state ($S_1$) corresponds to the nodeless Frenkel ground state, or local
exciton ground state (LEGS) \cite{Barford17,Barford12a} at $\theta=0^\circ$, while the second excited
state ($S_2$) exhibits a central node, in line with the ascending solutions of
a $J$-type Frenkel model \cite{Spano14,Binder2014}.
  At $\theta=90^\circ$, though, $S_1$ exhibits a
conjugation break, such that the $S_1$ and $S_2$ densities are very similar.
The transition between these limiting cases happens between 140$^{\circ}$-160$^{\circ}$ \cite{SI}.
It is this transition which underlies the dynamics observed in Fig.~\ref{figure2}.

Fig.\ \ref{figure4} illustrates the key steps of the dynamics in terms of the
coupled dynamics of the two lowest adiabatic excitonic states
\cite{Panda2013,Faraday2013}. From this viewpoint, the initial state is a left
or right localized coherent superposition, $\vert \psi_{\rm L} \rangle =
{1}/{\sqrt{2}}( \vert \psi_{S_1} \rangle + \vert \psi_{S_2} \rangle )$ or
$\vert \psi_{\rm R} \rangle = {1}/{\sqrt{2}}(\vert \psi_{S_1} \rangle - \vert
\psi_{S_2} \rangle)$. The initial, localized state is stabilized by trapping {\em via} 
    the high-frequency modes (Fig.\ \ref{figure4}a), with typical
    stabilization energies around $\Delta E_{\rm tr} \sim 0.5$ eV.
    The trapped state is, however, not fully relaxed, and sustained high-frequency oscillations
    are observed for hundreds of femtoseconds, in line with
    experiment \cite{Song2015}. 
    As long as the conjugation break persists, both $S_1$ and $S_2$ states
    exhibit density away from the junction between the left/right segments.
    The situation changes markedly as the system planarizes,
    such that the $S_1$ state turns into the nodeless LEGS on the full lattice.
    During the downhill dynamics on effective, dressed torsional 
    PESs (Fig.\ \ref{figure4}b), excess energy is absorbed by the external bath modes. 
   This picture is in line
    with experimental observations indicating correlated exciton relaxation
    driven by torsional motion \cite{Blank2008}.

Our analysis supports the notion of spectroscopic units \cite{Grage03,Beenken04}, in the sense of 
conformational subunits whose location and spatial extension dynamically evolves as a function of
conformational fluctuations. The present simulations, at $T=0K$, provide a snapshot
of such a torsional fluctuation event guiding coherent exciton migration.
At higher temperatures, thermally induced torsional fluctuations
will lead to multiple exciton migration pathways evolving under the influence of disorder. The high-frequency modes, by
contrast, remain unaffected by temperature, preserving the
polaronic character of the dressed exciton. The resulting picture
corresponds to a Holstein polaron driven by fluctuations.

The present analysis underscores the quasi-particle nature of the
exciton and the key role of exciton trapping, in contrast to
current interpretation, e.g., in Ref.\ [\citenum{Barford17}].
However, our results also suggest that the observed dynamics falls into a non-universal parameter regime where exciton
binding energies can differ markedly depending on exciton-phonon
coupling strength.

By capturing the elusive exciton migration event, the present analysis
offers a consistent interpretation of various experimental findings,
including the observation of ultrafast polarization anisotropy decay \cite{Grage03b},
the correlated ultrafast dynamics and large Stokes shift in conjugated polymers featuring torsional modes \cite{Blank2008}, the interleaved nature of
torsional relaxation and excitation energy transfer \cite{Sundstrom06}, and 
the persistence of
coherent high-frequency motions in photoexcited conjugated polymers \cite{Song2015}.
Thermal effects and the important role of aggregation and aggregation-induced planarization 
in P3HT \cite{Hildner2016} can be included in the present model and will be
addressed in forthcoming work. 

This work was supported by the Deutsche Forschungsgemeinschaft (DFG grant BU-1032-2).

\end{document}